\documentclass[11pt,english]{article}
\pdfoutput=1
\usepackage{jheppub,empheq}
\usepackage{hyperref}
\usepackage{simplewick}
\usepackage[title]{appendix}
\usepackage{verbatim}
\usepackage{subfigure}
\usepackage{graphics, color}
\usepackage{graphicx}
\usepackage{mathabx}
\usepackage{amsmath,mathtools}
\usepackage{amssymb}
\usepackage{amsthm}
\usepackage{appendix}
\usepackage{array}
\usepackage{tabularx}
\usepackage{amsfonts}
\usepackage{lipsum}   
\usepackage[usenames,dvipsnames]{xcolor}
\usepackage[normalem]{ulem}
\usepackage{braket}
\usepackage{titlesec}

\newcommand{\zb}{\bar{z}}
\newcommand{\Zb}{\bar{Z}}
\newcommand{\disc}{\text{disc}}
\newcommand{\hb}{\bar{h}}
\newcommand{\Tr}{\text{Tr}}

\title{Analyticity of replica correlators and Modular ETH}
\author[]{Milind Shyani}

\affiliation[]{Stanford Institute for Theoretical Physics, Department of Physics, Stanford University, CA 94305, USA.}

\emailAdd{shyani@stanford.edu}

\abstract{We study the two point correlation function of a local operator on an $n$-sheeted replica manifold corresponding to the half-space in the vacuum state of a conformal field theory. In analogy with the inverse Laplace transform, we define the Renyi transform of this correlation function, which is a function of one complex variable $w$, dual to the Renyi parameter $n$. Inspired by the inversion formula of Caron-Huot, we argue that if the Renyi transform $f(w)$ has bounded behavior at infinity in the complex $w$ plane, the discontinuity of the Renyi transform disc$f(w)$ provides the unique analytic continuation in $n$ of the original replica correlation function. We check our formula by explicitly calculating the Renyi transform of a particular replica correlator in a large $N$ holographic CFT$_d$ in dimensions $d>2$.

We also discover that the discontinuity of the Renyi transform is related to the matrix element of local operators between two distinct eigenstates of the modular Hamiltonian. We calculate the Renyi transform in $2d$ conformal field theories, and use it to extract the off-diagonal elements of (modular) ETH. We argue that in $2d$, this is equivalent to the off-diagonal OPE coefficients of a CFT and show that our technique exactly reproduces recent results in the literature.}

\newcommand{\beq}{\begin{equation}}
\newcommand{\eeq}{\end{equation}}
\newcommand{\beqn}{\begin{eqnarray}}
\newcommand{\eeqn}{\end{eqnarray}}

\begin{document}

\maketitle
\parskip=12pt

\section{Introduction}
Over the last few decades, it has become increasingly clear that entanglement entropy is an extremely useful quantity. It plays a crucial role in understanding a diverse set of phenomena ranging from topological order \cite{Kitaev:2005dm,Levin2006DetectingTO} to quantum gravity \cite{Ryu:2006bv,Hubeny:2007xt,Lewkowycz:2013nqa,Jafferis:2015del,Faulkner:2013ica,Dong:2016eik}. A standard measure of entanglement of a sub-region $A$ in a quantum state $\ket{\psi}$ is given by the von Neumann entropy,
\begin{align}
    S = - \Tr \rho \log \rho, \qquad \rho = \Tr_{\bar A} \ket{\psi} \bra{\psi},
\end{align}
where $\rho$ is the density matrix. The quantity $K \equiv -\frac{\log \rho}{2\pi}$ is known as the modular Hamiltonian. The von Neumann entropy is an extremely difficult quantity to compute in any interacting quantum system. 

For certain special shapes of the region $A$ and states $\ket{\psi}$ however, we can make progress \cite{doi:10.1063/1.522898,Casini:2007bt,Casini:2017roe}. In particular, the entanglement entropy of a quantum field theory associated to the half space has been widely studied in the literature \cite{Calabrese:2009qy,Faulkner:2013yia,Casini:2011kv}, and will serve as our starting point. The standard way to do this for the vacuum state is to calculate the $n$-th Renyi entropy by evaluating the path integral on the $n$-sheeted replica manifold,
\begin{align}
    \mathcal Z_n \equiv \Tr \rho^n.
\end{align}
The entanglement entropy can then be obtained by analytically continuing $n\rightarrow1$, and taking the derivative at $n=1$. It is widely believed that this analytic continuation is valid as long as the density of states grows sub-exponentially in the energy \cite{Rangamani:2016dms,Headrick:2019eth}. 

In this note we would like to study the two point correlation function of scalar operators $O(\vec x)$ on the $n$-sheeted replica manifold,
\begin{align}
   F_n(s,x) \equiv \Tr \left( \rho^{n} O(\vec x) \rho^{i \frac{s}{2\pi} } O(\vec x) \rho^{-i \frac{s}{2\pi}} \right). \label{mfir}
\end{align}
Such correlators arise while calculating entanglement entropy of excited states and/or shape deformations of the vacuum on the half plane, which have played an important role in recent developments \cite{Faulkner:2015csl,Bianchi:2016xvf,Belin:2018juv,Faulkner:2016mzt,Balakrishnan:2017bjg,Lewkowycz:2014jia,Faulkner:2014jva}. For example, in field theories with action deformed by $\delta I = - g \int O$, the perturbative expansion of $\log \mathcal Z_n$ is given by integrals of two (and higher) point correlators like (\ref{mfir}) \cite{Faulkner:2014jva,Lewkowycz:2014jia}. It is thus important to study the general properties of replica correlators such as (\ref{mfir}).

We study the analytic properties of $F_n(s,x)$ and provide the unique analytic continuation that is valid for non-integers value of $n$. The question regarding analytic continuation in the replica trick has been addressed before for several specific cases \cite{Cardy:2007mb,CastroAlvaredo:2008kh} where significant progress was reported, but our aim here is slightly different. We would like to develop a general technique that could, in principle, be used to address the question of analytic continuation in a wider class of examples. In this process, we discover multiple interesting identities. 

Our main motivation comes from Caron-Huot's revival of dispersion relations in his work on the inversion formula \cite{Caron-Huot:2017vep,Simmons-Duffin:2017nub}. Dispersion relations utilise the fact that an analytic function which is bounded at infinity is rather tightly constrained. For example, the standard Kramers-Kronig relations \cite{Jackson:100964} can be used to obtain the real part of an analytic function from its imaginary part and vice versa. However the analytic properties of the Renyi correlator $F_n(s,x)$ in the complex $n$ plane, as defined in (\ref{mfir}) are not obvious. 

To that end, inspired by the work of Calabrese and Lefevre \cite{Calabrese:2009qy} we define a new quantity, the Renyi transform $f(w,s,x)$ of the replica correlator $F_n(s,x)$. The Renyi transform is quite similar to the resolvent of the density matrix $\rho$ that has been used extensively in recent work \cite{Ugajin:2018rwd,Penington:2019kki}. In their work, Calabrese and Lefevre define the Renyi transform of the $n$-th replica partition function as, 
\begin{align}
f_{CL}(w) \equiv \sum_{n=1}^\infty   \Tr \rho^n \, w^n,
\end{align}
and show that the discontinuity of $f_{CL}(w)$ in the complex $w$ plane can be used to extract the distribution of eigenvalues of the modular Hamiltonian. Similarly, we define the Renyi transform $f(w,s,x)$ of the replica correlator,
\begin{align}
    f(w,s,x) \equiv \sum_{n=1}^\infty F_n(s,x) w^n. \label{def1}
\end{align}
This quantity has several benefits over $F_n(s,x)$. The first being that the analytic properties of $f(w,s,x)$ in the complex $w$ plane are far easier to understand than the analytic properties of $F_n(s,x)$ in the complex $n$ plane, as we argue in sections \ref{main} and \ref{prop}. In particular, we can show that if $f(w,s,x)/w$ is bounded at infinity in the complex $w$ plane, the discontinuity of the Renyi transform in the $w$ plane provides the unique analytic continuation for $F_n(s,x)$ for $\text{Re } n \geq 1$,
\begin{align}
F_n(s,x) = \frac{1}{2\pi i } \!\!\!\!\!\! \int\displaylimits_{w=e^{2\pi E_{min}}}^{\infty} \!\!\!\!\!\!\!\! dw  \, \, \frac{\disc \, f(w,s,x)}{w^{n+1}}, \label{sec}
\end{align}
where $E_{min}$ is the minimum eigenvalue of the modular Hamiltonian. In section \ref{cont} we show that, for nonzero $s$ and $x$, a sub-exponential density of states $\mu(E_n)$ and a sub-exponential growth of the three point function $\braket{E_n|O|E_m}$ in the modular energies $E_{n}$ and $E_m$ provide the necessary and sufficient set of criteria for $f(w,s,x)/w$ to be bounded.

Just like in \cite{Calabrese:2009qy}, where the discontinuity of $f_{CL}(w)$ in the complex $w$ plane was found to give the distribution of modular eigenvalues, we discover an important identity that relates the discontinuity of our Renyi transform and the matrix element $\braket{E_n|O|E_m}$,
\begin{align}
   \disc f\left(w=e^{2\pi E},\delta E,x\right)  = 2\pi i \mu(E) \mu(E+\delta E)|\braket{E |O(\vec x)|E + \delta E}|^2. \label{thir}
\end{align}
We believe this is a non-trivial result that provides a new method to extract the off-diagonal elements in the ETH ansatz \cite{shredder} in arbitrary dimensions. The quantity $\braket{E |O(\vec x)|E + \delta E}$ is the matrix element of a local operator in the eigenbasis of the modular Hamiltonian, and we call it the off-diagonal element of modular ETH. In conformal field theories this relation can be used to extract the Heavy-Light-$\widetilde{\text{Heavy}}$ OPE coefficients \cite{Fitzpatrick:2015zha} on Hyperbolic space, which are related to the thermalisation properties of Hyperbolic blackholes \cite{Casini:2011kv,Emparan_1999}. 

We perform some explicit calculations in the rest of the paper. In section \ref{2dcft}, we focus on $2d$ CFTs and explicitly compute the Renyi transform $f(w,s,x)$ of the replica correlator (\ref{mfir}) on the half plane. In section \ref{311}, we evaluate $\disc f(w,s,x)$ and compute the off-diagonal elements of modular ETH at asymptotically large energies. We find that they match exactly with the results on the averaged off-diagonal OPE coefficients $\overline{|C_{E_nOE_m}|^2}$ in any modular invariant $2d$ CFT \cite{Brehm:2018ipf,Romero-Bermudez:2018dim} at asymptotically large energies. We argue that the results necessarily have to match due to conformal invariance. We also confirm that $f(w,s,x)/w$ is bounded at infinity in the complex $w$ plane in section \ref{312}. In section \ref{holo}, we explicitly calculate the Renyi transform of the following  replica correlator,
\begin{align}
     \tilde F_n(x) \equiv \Tr \left( \rho^{n/2} O(\vec x) \rho^{n/2 } O(\vec x) \right),
\end{align}
in the context of large $N$ holographic CFT$_{d}$ for $d>2$ and find that (\ref{sec}) reproduces the right answer. The results of section \ref{2dcft} and \ref{holo} provide non-trivial checks of our results in equation (\ref{thir}) and (\ref{sec}) respectively. We conclude with possible generalisations and several future directions. 

\textit{Note added} -- After this work was completed an extremely interesting paper \cite{DHoker:2020bcv} appeared on the $arxiv$ that obtains the von Neumann entropy by analytically continuing a generating function $G(z,\rho)$ in a conjugate variable $z$. Although our end goals are different, I believe our approaches are morally similar. Indeed, to make the analogy precise, I obtain the replica correlator $F_n(s,x)$ by analytically continuing a generating function $f(w,s,x)$ in a conjugate variable $w$.


\section{Renyi Transform}
In this section we derive the central results of this paper. In section \ref{def} we define the Renyi transform for an arbitrary modular Hamiltonian and discuss issues regarding analytic continuation. In section \ref{prop} we specialise to the case of vacuum modular Hamiltonian and discuss the analytic properties of the Renyi transform. 
In section \ref{cont} we use the knowledge of the analytic structure from the previous subsections and write down the unique analytic continuation in $n$ of the replica correlator (\ref{main}). 



\subsection{Definition} \label{def}
Consider the following two point function of two scalar operators mentioned in the introduction,
\begin{align}
     F_n(s,x) & = \Tr \left( \rho^{n} O(\vec x) \rho^{i \frac{s}{2\pi} } O(\vec x) \rho^{-i \frac{s}{2\pi}} \right),  \label{main}
\end{align}
where $\rho = \Tr_{\bar{A}} \ket{\psi} \bra{\psi}$ is the reduced density matrix corresponding to some region $A$, $\vec x \in A$, and $n$ is a positive integer. We will call such two-point functions replica correlators. We want to find an analytic continuation for $F_n(s,x)$ that is valid for non-integer values of $n$.

Motivated by the work of Calabrese and Lefevre \cite{Calabrese_2008}, we first define \textit{the Renyi transform} of the replica correlator $F_n(s,x)$,
\begin{align}
    f(w,s,x) \equiv \sum_{n=1}^{\infty} F_n(s,x) w^n \label{ren}.
\end{align}
This expression can be inverted by the residue theorem to obtain,
\begin{align}
    F_n(s,x) = \frac{1}{2 \pi i }\oint_\mathcal{C} \frac{f(w,s,x)}{w^{n+1}}, \label{first}
\end{align}
where $\mathcal{C}$ is a tiny contour around $w=0$ in the complex $w$ plane as shown in figure \ref{wplane}. Equation (\ref{first}) makes sense only for integer $n$ since $w^{n+1}$ is multi-valued in the complex $w$ plane otherwise. If we are interested in finding an analytic function in $n$, we need to do better. In the following sections, we use the methods recently popularised by Caron-Huot in deriving the inversion formula \cite{Caron-Huot:2017vep,Simmons-Duffin:2017nub} to find the correct analytic continuation. 

Note that the form of the correlator in (\ref{main}) is chosen for simplicity. Our methods can be easily generalised for the replica correlator,
\begin{align}
     F_n(s,x_i,k) & = \Tr \left( \rho^{n(1-k)} O(\vec x_1) \rho^{kn}  \rho^{i \frac{s}{2\pi} } O(\vec x_2) \rho^{-i \frac{s}{2\pi}} \right), \qquad 1>k>0. \label{genrep}
\end{align}
This corresponds to the case when the operators are inserted on different replicas, and different spacetime positions.


\subsection{Analytic structure and relation to modular ETH} \label{prop}
Let us first analyse the analytic properties of $F_n(s,x)$ in the complex $n$ plane. Rewriting (\ref{main}) we have,
\begin{align}
    F_n(s,x) = \sum_{p,m}e^{-2 \pi n E_p  - i s(E_p-E_m)} |\braket{E_p|O(x)|E_m}|^2, \label{porid}
\end{align}
where $\rho = e^{-2\pi K}$. $K$ is the modular Hamiltonian, and $E_p$ and $E_m$ are its eigenvalues in the corresponding states. Since we will be working with infinite dimensional systems, the spectrum of the modular Hamiltonian is continuous and (\ref{porid}) becomes,
\begin{align}
    F_n(s,x) =\int_{E_{min}}^\infty dE_p \, \mu(E_p) \int_{E_{min}}^\infty dE_m \, \mu(E_m) e^{-2 \pi n E_p  - i s(E_p-E_m)} |\braket{E_p|O(\vec x)|E_m}|^2. \label{pori}
\end{align} 
where $\mu(E)$ represents the density of states. We will assume through out the entirety of this paper that the spectrum of the modular Hamiltonian is real, and is bounded from below by $E_{min}$. However, we shall not impose any UV cutoff on the modular energies $E_i$. Thus the replica correlator $F_n(s,x)$, given in (\ref{porid}) or (\ref{pori}), is not necessarily analytic in $n$, since it is given by an \textit{infinite} sum of terms (that are although analytic individually).

On the other hand, the analytic properties of the Renyi transform are easier to understand. Using (\ref{pori}) in (\ref{ren}) we have,
\begin{align}
    f(w,s,x) = \int_{E_{min}}^\infty dE_p \, \mu(E_p) \int_{E_{min}}^\infty dE_m \, \mu(E_m) \frac{w}{e^{2\pi E_p}-w}  e^{-i s (E_p - E_m)} |\braket{E_p|O(\vec x)|E_m}|^2. \label{intf}
\end{align}
From the integrand of this equation, we find that the Renyi transform $f(w,s,x)$ has a continuous density of poles in the complex-$w$ plane on the real axis  starting from $w=e^{2\pi E_{min}}$ to $\infty$.  The integrand is analytic everywhere else in the complex $w$ plane. We will see in the following sections that the integrals in (\ref{intf}) are convergent under the assumptions of sub-exponential growth in energy of the density of states $\mu(E_p)$ and matrix elements $\braket{E_p|O|E_m}$. Thus $f(w,s,x)$, just like the integrand, is analytic everywhere in the complex-$w$ plane except for the continuous density of poles on the real axis.

\begin{figure}
    \centering
    \includegraphics[scale=0.9]{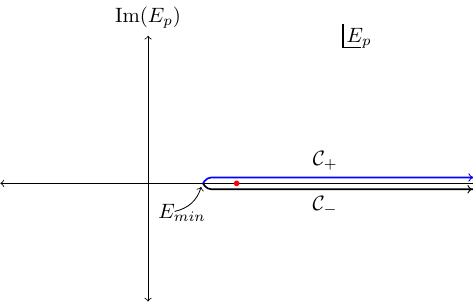}
    \hspace{0.4cm}
    \includegraphics[scale=0.9]{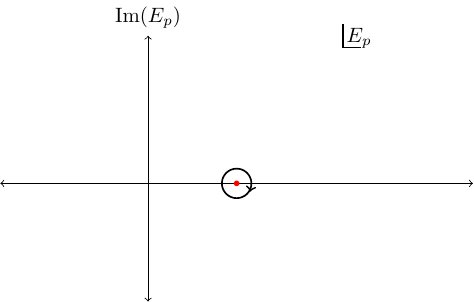}
    \caption{The integrand in (\ref{intf}) has a pole (red dot) in the complex $E_p$ plane at $E_p=\frac{\log(w)}{2\pi}$. The Renyi transform $f(w\pm i\epsilon,s,x)$ can be obtained by evaluating the $E_p$ integral on the contours $\mathcal C_{\pm}$ respectively. Although it is impossible to evaluate them individually, the difference (\ref{disc}) is simply given by the residue at the pole as shown in the right panel.}
    \label{eplane}
\end{figure}

To understand this better let us consider $f(w \pm i\epsilon, s,x)$ for some $\epsilon>0$ and $w \geq e^{2\pi E_{min}}$. Figure \ref{eplane} shows the analytic structure of the integrand in the complex $E_p$ plane. $f(w + i\epsilon, s,x)$ can be evaluated by integrating $E_p$ in (\ref{intf}) over the contour $\mathcal C_{+}$, and similarly for the other sign. The integral in (\ref{intf}) is a formal one that we cannot actually evaluate since we have no information about the density of states $\mu(E)$ or the matrix elements $\braket{E |O(\vec x)|E + \delta E}$. But consider the following quantity,
\begin{align}
    \disc f(w,s,x) \, \equiv f(w+i \epsilon,s,x ) - f(w - i \epsilon,s,x). \label{disc}
\end{align}
We call the left hand side the discontinuity of $f(w,s,x)$. The benefit of $\disc f$ is that it helps us evaluate the $E_p$ integral. As shown in figure \ref{eplane}, the difference of the two terms is just given by the residue at $E_p = \frac{\log w}{2\pi}$. To be precise we obtain,
\begin{align}
    \disc f\left(w=e^{2\pi E_p},s,x\right) = 2\pi i \mu(E_p) \int dE_m \, \mu(E_m)e^{-i s (E_p - E_m)}  |\braket{E_p|O(\vec x)|E_m}|^2.
\end{align}
Fourier transforming with respect to $s$,
\begin{align}
    \disc f\left(w=e^{2\pi E_p},\delta E,x\right) & = 2\pi i \mu(E_p) \int ds \int dE_m \, \mu(E_m)e^{-i s (E_p - E_m) - i s \delta E}  |\braket{E_p|O(\vec x)|E_m}|^2, \nonumber \\
    & =  2\pi i \mu(E_p) \int dE_m \, \delta(E_p -E_m+\delta E) \mu(E_m) |\braket{E_p|O(\vec x)|E_m}|^2,
\end{align}
we obtain the final form,
\begin{align}
\disc f\left(w=e^{2\pi E},\delta E,x\right)  = 2\pi i \mu(E) \mu(E+\delta E) |\braket{E |O(\vec x)|E + \delta E}|^2. \label{discmeth}
\end{align}
The above equation is one of the main results of this paper. It relates the discontinuity of the Renyi transform with the off-diagonal elements of local operators in modular energy eigenstates. We believe that this is a non-trivial result that could be used to find the off-diagonal piece in the Eigenstate Thermalisation Hypothesis (ETH) ansatz \cite{shredder}, applied to the eigenbasis of modular Hamiltonians. We refer to the right hand side of (\ref{discmeth}) as the off-diagonal piece of modular ETH. 

As mentioned in the introduction, in conformal field theories the elements of modular ETH give direct access to certain OPE coefficients of CFTs on hyperbolic space. This is morally similar to the relationship between the off-diagonal piece of ordinary ETH and OPE coefficients of CFTs on flat space \cite{Lashkari:2016vgj}. The $d$-dimensional $n$-sheeted cover of $\mathbb R^d$ branched at the boundary of the half space $x=0$ is given by,
\begin{align}
    ds^2 = x^2 d\theta^2 + dx^2 + dy_i^2, \qquad i=1,\ldots ,d-2, \label{hhe}
\end{align}
where $x>0, \, \theta = \theta + 2\pi n$ and $y_i \in (-\infty,\infty)$. This can be conformally mapped to $S_1^\beta \times \mathbb H_{d-1}$,
\begin{align}
    ds^2 =  d\theta^2 + \frac{dx^2 + dy_i^2}{x^2} . 
\end{align}
The modular Hamiltonian that generates translations in $\theta$ is conformally equivalent to the ordinary Hamiltonian that generates time translations on, 
\begin{align}
    ds^2 = -dt^2 + \frac{dx^2 + dy_i^2}{x^2}. 
\end{align} 
We thus find that in a conformal field theory the matrix element of an operator $O(\vec x)$ between different modular eigenstates $\braket{E |O(\vec x)|E + \delta E}$ are conformally related to the matrix elements of $O(\vec x)$ between the eigenstates of the ordinary Hamiltonian on Hyperbolic space $\mathbb H_{d-1}$. For large $N$ holographic CFTs these matrix elements are directly related to thermalisation properties of black holes with hyperbolic horizons \cite{Emparan_1999,Casini:2011kv}. Equation (\ref{discmeth}) thus provides us a new tool for investigating thermalisation in interacting field theories. In section \ref{2dcft}, we explicitly evaluate the Renyi transform in $2d$ CFTs and obtain these matrix coefficients at asymptotically large energies.

The other piece of information that we obtain from $(\ref{discmeth})$ is that the Renyi transform $f(w,s,x)$ has a branch cut in the complex-$w$ plane starting from $w=e^{2\pi E_{min}}$ to $\infty$. In other words the continuous density of poles mentioned earlier coalesce into a branch cut. Having obtained this information we turn back to equation (\ref{first}) in the following section to obtain an analytic formula for $F_n(s,x)$ by deforming the integration contour $\mathcal C$ appropriately.


\subsection{Analytic continuation} \label{cont}
Recall our formula (\ref{first}) from the previous section,
\begin{figure}
    \centering
    \includegraphics[scale=0.5]{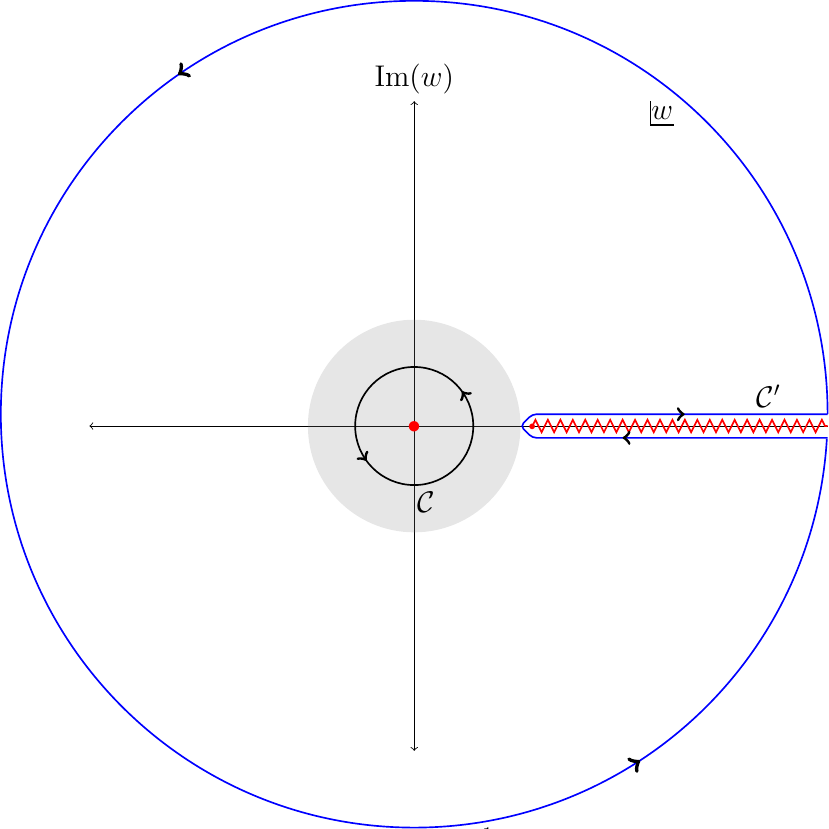}
    \caption{The function $f(w,s,x)$ is analytic in $w$ in the shaded disc, but has a branch cut along the real axis starting at $w = e^{2\pi E_{min}}$ (red line). The contour $\mathcal C$ that goes around the pole of $f(w,s,x)/w^{1+n}$ at the origin, can be deformed in to $\mathcal C'$. This contour wraps the branch cut starting from $e^{2\pi E_{min}}$ to $\infty$. If $f(w,s,x)/w$ is bounded at infinity, the arcs at infinity can be dropped to obtain (\ref{contf}).}
    \label{wplane}
\end{figure}
\begin{align}
    F_n(s,x) = \frac{1}{2 \pi i }\oint_\mathcal{C} dw \frac{f(w,s,x)}{w^{n+1}}.
\end{align}
As shown in figure \ref{wplane} we can deform the integration contour $\mathcal C$ into $\mathcal C^\prime$ by wrapping it around the branch cut. If the Renyi transform  satisfies the following property,
\begin{align}
    \lim_{w\rightarrow\infty}\frac{f(w,s,x)}{w} < \infty, \label{bounded}
\end{align}
we can drop the arcs at infinity to obtain the convergent integral,
\begin{align}
    F_n(s,x) = \frac{1}{2\pi i } \!\!\!\!\!\! \int\displaylimits_{w=e^{2\pi E_{min}}}^{\infty} \!\!\!\!\!\!\!\! \frac{dw}{w}  \, \, \frac{\disc \, f(w,s,x)}{w^n}, \label{contf}
\end{align}
for all $ n > 1$. If $f(w,s,x)/w$ vanishes as $w\rightarrow\infty$ in (\ref{bounded}), instead of just being bounded, then (\ref{contf}) would also hold for $n=1$.

This is one of the main results of our paper. Equation (\ref{contf}) is analytic in $n$ and is valid for all $\text{Re} \hspace{0.05cm} n>1$. Below we make an argument that a sub-hagedorn density of states is a necessary and sufficient condition for (\ref{bounded}). In certain cases, like $2d$ CFTs as shown in section \ref{312}, we can argue that that $f(w,s,x)/w$ vanishes as $w \rightarrow \infty$. By Carlson's theorem \cite{carlson} then, (\ref{contf}) provides the unique analytic continuation for all $\text{Re} \hspace{0.05cm} n \geq 1$. However a limitation of this formula is that it is valid only when $s\neq 0$ and $x \neq 0$. This is because (\ref{bounded}) is unbounded when $s\rightarrow 0$ or $x \rightarrow 0$. Recall from (\ref{main}) that $s=0$ corresponds to a coincident point singularity inside the correlator, while $x=0$ corresponds to approaching the entanglement cut.\footnote{For the more general replica correlator (\ref{genrep}) it can be similarly shown that $f(w,s,x)/w^n$ is bounded as long as there are no coincident point singularities $s \neq 0  \lor x_1 \neq x_2 \lor 1>k>0$, and the operators are inserted away from the cut $x_1 \neq 0 \land x_2 \neq 0$.}

In the remainder of this section we argue that a sub-hagedorn density of states is a necessary and sufficient condition for (\ref{bounded}) to hold true, as long as there are no coincident point singularities and the operators do not approach the entanglement cut. In later sections we explicitly compute the Renyi transform and find that $f(w,s,x)/w$ vanishes at complex infinity in the $w$ plane. Recall that the Renyi transform is given by (\ref{intf}),
\begin{align}
     f(w,s,x) =\int dE_p \mu(E_p) \int dE_m \mu(E_m) \frac{ w e^{-i s (E_p - E_m)} }{e^{2\pi E_p }-w}   |\braket{E_p|O(x)|E_m}|^2. \label{fir}
\end{align}
We are interested in the limit,
\begin{align}
    \lim_{w \rightarrow \infty} \frac{f(w,s,x)}{w} = \lim_{w \rightarrow \infty} \frac{1}{w } \int_{E_{min}}^\infty dE_p \mu(E_p) \int_{E_{min}}^\infty dE_m \mu(E_m) \frac{w e^{-i s (E_p - E_m)} }{e^{2\pi E_p }-w}   |\braket{E_p|O(\vec x)|E_m}|^2. \label{lim}
\end{align}
It is obvious that the above is bounded if we take the limit $w\rightarrow \infty$ first. However, the range of the integral is unbounded, and we need to evaluate the integral before we take the limit. In other words, the limit and the integral do not necessarily commute. 

We will try to evaluate the integral for large but finite $w$. Let us first take the fourier transform in $s$ to make the integrals more tractable. This is a harmless transformation that can be made on both sides of all our results, including (\ref{contf}). We get,
\begin{align}
      \int ds \frac{f(w,s,x)}{w} e^{-i s \delta E} & =\frac{1}{w } \int_{E_{min}}^\infty dE_p \mu(E_p) \int_{E_{min}}^\infty dE_m \mu(E_m) \int ds \frac{w e^{-i s (E_p - E_m + \delta E)} }{e^{2\pi E_p }-w}   |\braket{E_p|O(x)|E_m}|^2. \nonumber \\
     \frac{f(w,\delta E,x)}{w} & = \int_{E_{min}}^\infty dE e^{-2\pi E }\left(\frac{\mu(E) \mu(E + \delta E) e^{2\pi E } \ }{e^{2\pi E }-w}   |\braket{E|O|E + \delta E}|^2 \right). \label{int}
\end{align}
In the last line we have separated the exponential to make the integral look like the Laplace transform in the variable $E$. It is a standard result that the Laplace transform is well defined if and only if the integrand grows sub-exponentially in the variable. The existence of the Laplace transform in $E$ however does not guarantee that the integral is bounded as we take $w \rightarrow \infty$. Thus the existence of the Laplace transform, \textit{a priori} is only necessary but not a sufficient criteria for boundedness (\ref{bounded}).

We first make the following assumptions for the existence of the integral. We assume that the density of states is sub-Hagedorn, 
\begin{align}
    \mu(E) \lesssim e^{c_0 E^{\gamma_0}}, \quad \gamma_0 < 1, \label{den}
\end{align}
and that the three point function $\braket{E|O|E+\delta E}$ grows at the most sub-exponentially in energy,  
\begin{align}
    \braket{E|O|E+\delta E} \lesssim e^{c_1 E^{\gamma_1}}, \quad \gamma_1 < 1 \label{denope}.
\end{align}
Here $c_i$ are some $E$ independent constants.

We will now show that this set of assumptions is also sufficient for (\ref{bounded}). We are interested in the integral,
\begin{align}
    \frac{f(w,\delta E,x)}{w} & = \int_{E_{min}}^\infty dE \frac{\mu(E)  \mu(E+\delta E)}{e^{2\pi E }-w}   |\braket{E|O|E + \delta E}|^2 .
\end{align}
Under the assumptions (\ref{den}) and (\ref{denope}), this integral can be bounded by,
\begin{align}
    \frac{f(w,\delta E,x)}{w} \leq I \equiv  \int_{E_{min}}^\infty \frac{dE}{e^{2\pi E} - w} e^{\tilde c E^\gamma} 
\end{align}
for some constant $\tilde c$ and $\gamma<1$. Making a variable change $e^{2\pi E} = \alpha |w|$, we obtain
\begin{align*}
    I & = \frac{1}{2 \pi |w|}\int_{\frac{m}{|w|}}^\infty \frac{d\alpha}{\alpha}\frac{1}{\alpha - e^{i \theta}} e^{c (\log(\alpha |w|))^\gamma} 
\end{align*}
where $c = \frac{\tilde c}{(2\pi)^\gamma}, m= e^{2\pi E_{min}}$ and $w \equiv |w| e^{i \theta}$. The arcs at complex infinity in the $w$ plane correspond to $|w|\rightarrow \infty$ for $\theta \in (0,2\pi)$. We first split the above integral into two,
\begin{align*}    
    I & = \frac{1}{2 \pi |w|}\int_{\frac{m}{|w|}}^{|w|-\epsilon} \frac{d\alpha}{\alpha}\frac{1}{\alpha - e^{i \theta}} e^{c (\log(\alpha |w|))^\gamma} + \frac{1}{2 \pi |w|}\int_{|w|+\epsilon}^\infty \frac{d\alpha}{\alpha}\frac{1}{\alpha - e^{i \theta}} e^{c (\log(\alpha |w|))^\gamma}, 
\end{align*}
as $\epsilon$ goes to zero. This would allow us to form the following inequality,
\begin{align}    
   I  & \lesssim \frac{e^{c \left(2\log|w|\right)^\gamma}}{2 \pi |w|}  \int_{\frac{m}{|w|}}^{|w|-\epsilon} \frac{d\alpha}{\alpha}\frac{1}{\alpha - e^{i \theta}}  + \frac{1}{2 \pi |w|}\int_{|w|+\epsilon}^{\infty} \frac{d\alpha}{\alpha}\frac{1}{\alpha - e^{i \theta}} e^{c (2\log(\alpha))^\gamma} \nonumber \\
     & \lesssim  \frac{e^{c \left(2\log |w|\right)^\gamma}}{2 \pi |w|} \log \left(\frac{m (|w|-e^{i \theta}-\epsilon)}{(|w|-\epsilon)(w-m)}\right)  + \frac{1}{2 \pi |w|}\int_{|w|+\epsilon}^{\infty} \frac{d\alpha}{\alpha}\frac{1}{\alpha - e^{i \theta}} e^{c (2\log(\alpha ))^\gamma}.
\end{align}
It can be checked that the first term vanishes as $w \rightarrow \infty$ as long as $\gamma$ is strictly smaller than one. We just need to ensure that the second term is bounded as well. It can be easily shown that the integral in the second term is convergent as long as $\gamma$ is strictly smaller than one. Thus all that remains to be shown is that the integral, 
\begin{align}
    \tilde{I}(w) \equiv \int_{|w|+\epsilon}^{\infty} \frac{d\alpha}{\alpha}\frac{1}{\alpha - e^{i \theta}} e^{c (2\log(\alpha ))^\gamma},
\end{align}
is at the most linear in $w$. This can be easily seen by using Leibniz's integral rule,
\begin{align}
    \frac{d \tilde I(w)}{d w} = - \frac{e^{c (2\log(\alpha ))^\gamma}}{\alpha(\alpha-e^{i \theta})} \bigg|_{\alpha=|w|+\epsilon} = - \frac{e^{c (2\log(|w| + \epsilon))^\gamma}}{(|w|+\epsilon)(|w|+\epsilon-e^{i \theta})}.
\end{align}
Since $\gamma$ is strictly smaller than one, the RHS is bounded by a constant. Consequently, $\tilde I(w)$ grows at most linearly in $w$. This argument could be possibly extended to argue that $\tilde I(w)$ grows strictly slower than $w$, but we shall not attempt it here. In summary, we have shown that for a subhagedorn density of states and sub-exponential growth of the three point function, 
\begin{align}
    \lim_{w \rightarrow \infty}\frac{f(w,\delta E, x)}{w} < \infty, \qquad \forall \delta E <\infty \land x \neq 0. \label{proof}
\end{align}

Before we end this section we would like to emphasize the following important caveat. If $\delta E \rightarrow \infty$ or $x=0$, then $c_1$ in (\ref{denope}) could in principle be unbounded and none of the above would be true. Thus (\ref{proof}) is valid only when $\delta E$ is finite and the operators are inserted away from the entanglement cut. In terms of the conjugate variable $s$, this means that $f(w,s,x)/w$ is bounded at complex infinity in the $w$ plane for all $s$ except $s=0$, and $x=0$. Recall from our original definition of the replica correlator (\ref{main}), that $s=0$ corresponds to a coincident point singularity. Our calculation for $2d$ CFTs in section \ref{312} exhibits the same physics, $f(w,s,x)/w$ vanishes at complex infinity in the $w$ plane except when $s=0$ or $x=0$.



\section{2d CFT} \label{2dcft}
We will now perform some explicit calculations to provide further evidence for our results from the previous sections. In this section, we calculate the Renyi transform $f(w,s,x)$ in a $2d$ CFT and find that $f(w,s,x)/w$ vanishes at infinity in the complex $w$ plane. We also calculate the discontinuity of $f(w,s,x)$ and obtain the off-diagonal elements of modular ETH. 

Consider the case of an interval $[u,v]$ in a two dimensional euclidean CFT. This can be viewed as a subsystem in an infinitely long one dimensional system in Lorentzian signature. We will follow the conventions of Calabrese and Cardy \cite{Calabrese:2009qy}, and use the complex coordinate $w=x+ i \tau$ and $\widebar w = x - i \tau$. Using the conformal map, 
\begin{align}
Z= \frac{w-u}{v-w}, \label{zw}
\end{align}
we can map this interval to the infinite half line $[0,\infty)$. 

\begin{figure}
    \centering
    \includegraphics[scale=0.62]{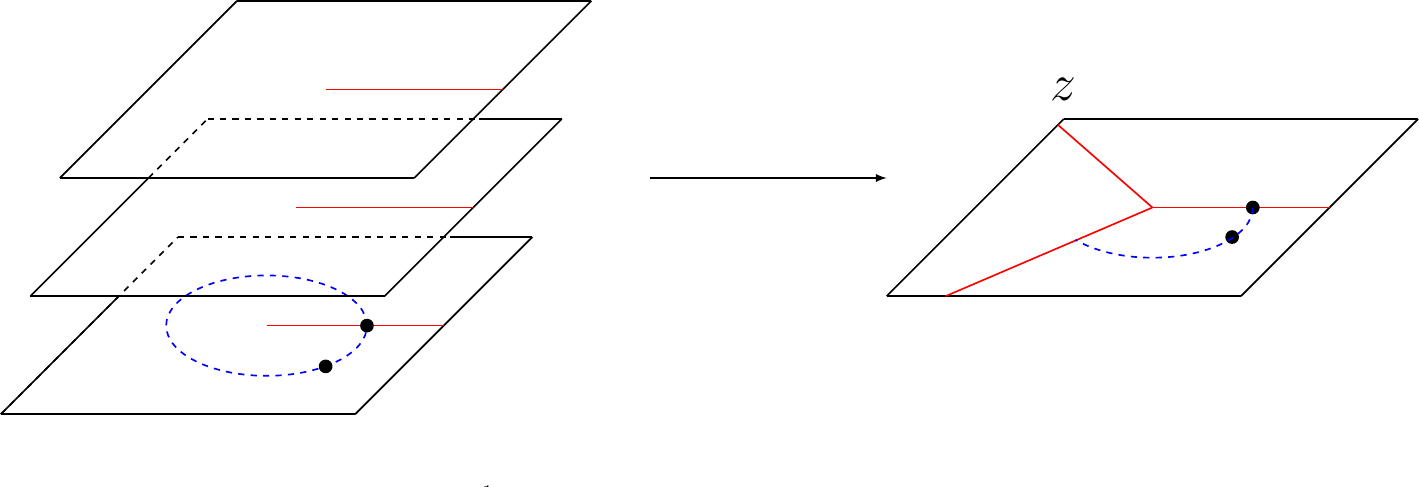}
    \caption{The coordinate transformation $z=Z^{1/n}$ maps the $n$-sheeted manifold to the complex plane. Here $n=3$.}
    \label{conf}
\end{figure}

The Renyi entropy of the vacuum state corresponding to the half line can be computed by evaluating the path integral on the $n$-sheeted Riemann surface $\mathcal M_2$ with branch points at $0$ and $\infty$. We are interested in calculating the two point function of scalar operators $F_n(w,x,s)$ on this background,
\begin{align}
    \braket{O(Z_1, \bar{Z}_1) O(Z_2, \bar{Z}_2)}_{\mathcal M_2}.
\end{align}
This is very hard to compute in a generic quantum field theory, but for a CFT we can uniformize the Riemann surface to the complex plane by the conformal map as shown in figure \ref{conf},
\begin{align}
    Z \rightarrow z= Z^{1/n}.
\end{align}
The two point function is given by,
\begin{align}
    \braket{O(Z_1,\Zb_1) O(Z_2,\Zb_2)}_{\mathcal M_2} &  = \left(\frac{dz_1}{dZ_1} \right)^h
    \left(\frac{d\zb_1}{d\Zb_1} \right)^{\hb} \left(\frac{dz_2}{dZ_2} \right)^h
    \left(\frac{d\zb_2}{d\Zb_2} \right)^{\hb} \braket{O(z_1,\zb_1) O(z_2,\zb_2)}. \label{2pn1}
\end{align}
    Let the points be inserted at $Z_1 = \bar Z_1 = x_1$ and $Z_2=x_2 e^{i \theta}, \bar Z_2=x_2 e^{-i \theta}$. Using the conformal two point function on the plane,
\begin{align}
    \braket{O(z_1,\zb_1) O(z_2,\zb_2)} = \frac{c_{h,\hb}}{\left(z_1- z_2 \right)^{2h}\left(\zb_1- \zb_2 \right)^{2\hb}}, \label{cft2}
\end{align}
the two point function on the replica manifold is given by,
\begin{align}
    \braket{O(Z_1,\Zb_1) O(Z_2,\Zb_2)}_{\mathcal M_2} & =  \frac{1}{n^{2\Delta} (x_1 x_2)^{\Delta\frac{n - 1}{n}}}  \frac{c_{h,\hb}}{\left(x_1^{1/n}  - x_2^{1/n} e^{i \theta/n} \right)^{2h}\left(x_1^{1/n} - x_2^{1/n} e^{-i \theta/n} \right)^{2\hb}}, \label{2pn}
\end{align}
where we have used $\Delta = h + \hb$, the conformal dimension of the operator $O$. Note that except for $n = 1 $, the replica correlator (\ref{2pn}) has a singularity as $x_1$ or $x_2$ approaches the branch point $x_1=0$ or $x_2=0$.

Recall from our definition of the replica correlator (\ref{main}) that $F_n(s,x)$ is the un-normalised two point function on the replica manifold. Since the CFT two point function (\ref{cft2}) is normalised, we have, 
\begin{align}
   \frac{F_n(Z_i,\bar Z_i)}{\text{Tr}(\rho^n)} = \braket{O(Z_1,\Zb_1) O(Z_2,\Zb_2)}.
\end{align}
We will now use the general result of Calabrese and Lefevre \cite{Calabrese_2008} for the Renyi partition function for 1D quantum systems close to criticality,
\begin{align}
    \text{Tr}(\rho^n) \equiv R_n = c_n e^{-b\left(n - \frac 1 n\right)}, \qquad b=2 \pi E_{min}, 
\end{align}
where $E_{min}$ is the minimum energy eigenvalue of the modular Hamiltonian. It was argued there that $c_n \equiv \boldsymbol c $ can be assumed to be independent of $n$. For the half line in a 2d CFT, it is a fact that $c_n$ is indeed independent of $n$ \cite{Calabrese:2009qy,Holzhey:1994we}. In the rest of the section we will also take $x_1 = x_2=x$. Putting this all together and analytically continuing to real time $\theta = - i s + \epsilon$ we finally obtain the replica correlator as defined in (\ref{main}), 
\begin{align}
   F_n(s,x) =  R_n \frac{x^{-2\Delta}}{n^{2\Delta}}  \frac{c_{h,\hb}}{  \left(1  - e^{ \frac{ i \epsilon + s}{n}} \right)^{\Delta} \left(1 -  e^{- \frac{ i \epsilon +  s}{n}} \right)^{\Delta}}.
\end{align}
Here $\epsilon$ is a positive infinitesimal number to ensure that the real time correlator is time ordered.

Having obtained the replica correlator we shall now explicitly calculate the Renyi transform from its definition (\ref{ren}),
\begin{align}
    f(w,s,x) & =  \sum_{n=1}^{\infty}  F_n w^n, \nonumber \\
    & = \frac{c_{h,\hb}}{x^{2\Delta} } \sum_{n=1}^\infty  \frac{R_n}{n^{2\Delta}}  \frac{w^n}{  \left(- 4 \sinh^2\left(\frac{s}{2n}  + i \epsilon\right) \right)^{\Delta}}.
\end{align}
In evaluating the sum we first take the fourier transform to obtain,
\begin{align}
    \int ds e^{-i s \delta E} f(w,s,x) \equiv f(w,\delta E,x)  =  \frac{c_{h,\hb}}{x^{2\Delta} \pi} \sum_{n=1}^\infty  \frac{R_n w^n}{n^{2\Delta}}   \int ds  \frac{ e^{-i s \delta E}}{  \left(- 4 \sinh^2\left(\frac{s}{2n}   + i \epsilon \right) \right)^{\Delta}}.
\end{align}
This is a standard integral which evaluates to,
\begin{align}
     f(w,\delta E,x) = \frac{c_{h,\hb}\boldsymbol c}{x^{2\Delta}} \sum_{n=1}^\infty  \frac{e^{-b\left(n- \frac{1}{n}\right)} w^n}{n^{2\Delta}}   \frac{n e^{ \pi n \delta E} \Gamma(\Delta - i n \delta E)\Gamma(\Delta + i n \delta E)}{\Gamma(2\Delta)} . \label{pre}
\end{align}
To evaluate the sum, let the Gamma functions have the following series expansion,
\begin{align}
    \Gamma\left(\Delta - i n \delta E \right)\Gamma\left(\Delta + i n \delta E \right) \equiv \sum_{j=0}^\infty a_j(\Delta) \left( i n \delta E\right)^j, \label{gam}
\end{align}
for some $\Delta$ dependent constants $a_j(\Delta)$. Expanding the exponential $e^{\frac{b}{n}}$ and using the expansion for the Gamma function we get,
\begin{align}
     f(w,\delta E,x) & = \frac{c_{h,\hb}\boldsymbol c}{x^{2\Delta} \Gamma(2\Delta) } \sum_{n=1}^\infty  \frac{(we^{-b  + \pi  \delta E })^n  }{n^{2\Delta-1}}  \sum_{k=0}^{\infty} \frac{b^k}{k!n^k}   \sum_{j=0}^{\infty} a_j (i n \delta E)^j .
\end{align}
The sum over $n$ can be evaluated using the definition of the Polylog function $\text{Li}_\nu(z)$,
\begin{align}
     f(w,\delta E,x) & =  N   \sum_{j=0}^\infty a_j (i  \delta E)^j \sum_{k=0}^\infty \frac{b^k}{k!}   \text{Li}_{2\Delta - 1 + k - j}(W), \label{2dansf}
\end{align}
where $N \equiv \frac{c_{h,\hb}\boldsymbol c}{x^{2\Delta} \pi \Gamma(2\Delta) }$ is a normalisation constant and $W \equiv we^{-b  + \pi  \delta E}$. This is the main result of this section. Although we do not have a closed form expression for the Renyi transform we will be able to extract the off-diagonal elements of modular ETH at sufficiently high energies using (\ref{discmeth}) in section \ref{311}, and show boundedness as $w \rightarrow \infty$ in section \ref{312}.

\subsection{Off-diagonal elements of modular ETH} \label{311}
We will now evaluate the discontinuity of $f(w,\delta E, x)$ to obtain the off-diagonal elements of the modular Hamiltonian. Using the discontinuity of the polylogarithm,
\begin{align}
    \text{disc Li}_{\nu}(z) \equiv   \text{Li}_\nu(z+ i \epsilon) - \text{Li}_\nu(z- i \epsilon)  = 2 \pi i \frac{\ln^{\nu-1} z}{\Gamma(\nu)}, \quad  \text{for} \quad z>1, \label{li}
\end{align}
in our expression for the Renyi transform (\ref{2dansf}) we find,
\begin{align}
    \disc f(w=e^{2\pi E},\delta E,x) & =  2 \pi i N \sum_{j=0}^\infty a_j (i \delta E)^j \sum_{k=0}^{\infty} \frac{b^k}{k!} \frac{\left(\log W\right)^{2\Delta - 2 + k -j}}{\Gamma(2\Delta - 1 + k - j)}. \nonumber
\end{align}
Here $W=e^{2\pi \left(E - E_{min}  + \delta E\right)}$, and the inequality in (\ref{li}) imposes,
\begin{align}
    E +  \delta E > E_{min} .
\end{align}
The sum over $k$ can be evaluated using the series expansion of the modified Bessel function $I_\nu(z)$ to obtain,
\begin{align}
     \disc f(w=e^{2\pi E},\delta E,x) & = 2 \pi i N \sum_{j=0}^\infty a_j (i \delta E)^j \left(\frac{\log W }{b}\right)^{\Delta-\frac j 2 - 2} I_{2\Delta - j - 2}(2\sqrt{b \log W}). \label{disc2d}
\end{align}
The sum over $j$ is difficult to evaluate, and we haven't been able to do it in closed form. However, we can make progress when the argument of the Bessel function is asymptotically large. $I_\nu(z)$ for large $z$, has the following well known expansion,
\begin{align}
    I_\nu(z) = \frac{1}{\sqrt{2\pi z}} \left[ e^z \left( 1 +  O\left(\frac 1 z\right)\right) + e^{-z + i \pi \nu  } \left( 1 +  O\left(\frac 1 z\right)\right) \right]. \label{serbel}
\end{align}
At asymptotically large energies, the argument $b \log W  $ in (\ref{disc2d}) is asymptotically large, and we can use (\ref{serbel}) to rewrite it as, 
\begin{align}
     \disc f(w=e^{2\pi E},\delta E,x)  = &  2 \pi i N \sum_{j=0}^\infty a_j (i \delta E)^j \left(\sqrt{\frac{\log W}{b}}\right)^{2\Delta -j -2} \frac{e^{2\sqrt{b \log W}}}{\sqrt{4\pi \sqrt{b \log W} }}. \nonumber 
\end{align}
The final sum over $j$ can be evaluated by resumming it into a Gamma function using the series expansion of (\ref{gam}). Putting it all together we finally obtain,
\begin{align}
\disc f(w=e^{2\pi E},\delta E,x) = &  \Gamma\left(\Delta + i \delta E \sqrt{\frac{E_{min}}{ E - E_{min}  +  \frac{\delta E}{2} }} \right)\Gamma\left(\Delta - i \delta E \sqrt{\frac{ E_{min}}{ E -E_{min}  +           \frac{\delta E}{2} }} \right) \nonumber \\
     & \times   \frac{N  i }{ \sqrt 2}\frac{\left( E -  E_{min} +  \frac{\delta E}{2}\right)^{\Delta - \frac 5 4}}{{ E_{min}^{\Delta - \frac 3 4 } }} e^{4 \pi \sqrt{ E_{min} \left(E -  E_{min}  +  \frac{\delta E}{2} \right)}}.  \label{2dansdisc}
\end{align}
Using this and our result (\ref{discmeth}) from the earlier section,
\begin{align}
  \text{disc} f(e^{2\pi E} =w,\delta E,x) =   2\pi i \mu(E) \mu(E+\delta E)|\braket{E|O(x)|E+ \delta E}|^2, \label{dmeth}
\end{align}
we can obtain the matrix element of a scalar operator $O(\vec x)$ in the eigenbasis of the modular Hamiltonian corresponding to the vacuum state on a half space in a $2d$ CFT.

Before we write down the explicit form for $\braket{E|O(x)|E+ \delta E}$, we would like to reinterpret this result using the manipulations mentioned previously near (\ref{hhe}). Recall that $O(x)$ is an operator inserted on the Euclidean $n$-sheeted $2d$ manifold,
\begin{align}
    ds^2 = x^2 d\theta^2 + dx^2, \label{hyp2da}
\end{align}
where $\theta \equiv \theta + 2\pi n$ and $x>0$. We can conformally map this manifold to $S_1^\beta \times \mathbb H_1 $,
\begin{align}
    ds^2 = d\theta^2+\frac{dx^2}{x^2}, \label{hyp2d}
\end{align}
where $\beta = 2\pi n$. Like before, we find that we can reinterpret the modular energy eigenstates $\ket{E}$ as the eigenstates of the ordinary Hamiltonian that generates time translations on the Hyperbolic line $\mathbb H_1$. The scalar operator, under the conformal transformation that relates (\ref{hyp2da}) and  (\ref{hyp2d}), transforms as, 
\begin{align}
    O(x) \rightarrow |x|^\Delta O(x), \label{sca}
\end{align}
Using (\ref{2dansdisc}), (\ref{dmeth}) and (\ref{sca}) we finally obtain the off-diagonal matrix elements of a scalar operator in the eigenstates of the ordinary Hamiltonian in a CFT on $S_1^\beta \times \mathbb H_1 $,
\begin{align}
    |\braket{E|O(x)|E+ \delta E}|^2  = & \frac{ \sqrt 2 c_{h,\hb}\boldsymbol c }{ E_{min}^{\Delta - \frac 3 4 } \Gamma(2\Delta)} \left| \Gamma\left(\Delta + i \delta E \sqrt{\frac{E_{min}}{ E - E_{min}  +  \frac{\delta E}{2} }} \right)\right|^2  \nonumber  \\
    & \times \left( E -  E_{min} +  \frac{\delta E}{2}\right)^{\Delta - \frac 5 4}\frac{e^{4 \pi \sqrt{ E_{min} \left(E -  E_{min}  +  \frac{\delta E}{2} \right)}}}{\mu(E)\mu(E+\delta E)}.   \label{2dans}
\end{align}
This is the main result of this subsection. 

Since we are in two dimensions we can perform one final manipulation. Note that the Hyperbolic line is just the real line $\mathbb R$ as seen by the coordinate transformation $z = \log x$,
\begin{align}
    ds^2 = d\theta^2 + dz^2,
\end{align}
for $z \in (-\infty,\infty)$. We find that the matrix element of the ordinary Hamiltonian (\ref{2dans}) on $S_1^\beta \times \mathbb H_1$ is identical to the matrix element of the ordinary Hamiltonian that generates translations on the ordinary cylinder.\footnote{We would like to thank Raghu Mahajan for discussions regarding this point.} Using ETH for a conformal field theory on flat space at asymptotically large energies, the matrix element can be written as \cite{Lashkari:2016vgj,Brehm:2018ipf},
\begin{align}
    |\braket{E|O|E + \delta E}|^2 \sim  \overline{|C_{E,O,E+\delta E}|^2}. \label{eth} 
\end{align}
The bar represents the average over all high energy states in an $O(1)$ microcanonical window and does not distinguish between primaries and descendants. Comparing (\ref{eth}) with (\ref{2dans}) we obtain for nonzero $\delta E$,
\small{
\begin{align}
    \overline{|C_{E,O,E+\delta E}|^2} =  \sqrt 2 c_{h,\hb}\boldsymbol c  \left|\Gamma\left(\Delta + i \delta E \sqrt{\frac{E_{min}}{ E - E_{min}  +  \frac{\delta E}{2} }} \right)\right|^2  \frac{\left( E -  E_{min} +  \frac{\delta E}{2}\right)^{\Delta - \frac 5 4}}{{ E_{min}^{\Delta - \frac 3 4 } \Gamma(2\Delta) }} \frac{e^{4 \pi \sqrt{ E_{min} \left(E -  E_{min}  +  \frac{\delta E}{2} \right)}}}{\mu(E) \mu(E+\delta E)}.
\end{align}}
Up to a prefactor that arises due to conventions involving $c_{h,\hb}$, this result is identical to the recent results of \cite{Brehm:2018ipf,Romero-Bermudez:2018dim,Kraus:2016nwo} about the off-diagonal OPE coefficients in any modular invariant $2d$ CFT on flat space.

\subsection{Bounding $f(w,\delta E,x)/w$} \label{312}
In this section we check that the Renyi transform has the right properties at infinity in the complex $w$ plane. In section \ref{cont} we showed that $f(w,s,x)/w$ is bounded by using the assumptions of sub-exponential growth. For $2d$ CFTs since we have the calculated the Renyi transform we will be able to explicitly check its behaviour as $w \rightarrow \infty$. We start with (\ref{pre}) that we reproduce here for the reader's convenience, 
\begin{align}
      f(w,\delta E,x) = \frac{c_{h,\hb}\boldsymbol c}{x^{2\Delta} \pi} \sum_{n=1}^\infty  \frac{e^{-b\left(n- \frac{1}{n}\right)} w^n}{n^{2\Delta}}   \frac{n e^{ \pi n \delta E} \Gamma(\Delta - i n \delta E)\Gamma(\Delta + i n \delta E)}{\Gamma(2\Delta)} . \label{prec}
\end{align}
For real $a$ and $b$ we have the following inequality,
\begin{align}
    \Gamma(a)^2 \geq \Gamma(a + i b) \Gamma(a - i b),
\end{align}
which gives,
\begin{align}
      f(w,\delta E,x) \leq \frac{c_{h,\hb}\boldsymbol c}{x^{2\Delta} \pi} \frac{\Gamma(\Delta)^2}{\Gamma(2\Delta)} \sum_{n=1}^\infty  \frac{e^{-b\left(n- \frac{1}{n}\right)} w^n}{n^{2\Delta-1}}    e^{ \pi n \delta E} .
\end{align}
If we define a function such that,
\begin{align}
   g(b) & =  \begin{cases} e^{b}  &\mbox{if } \quad b\geq 0 \\
1 & \mbox{if } \quad b \leq 0 \end{cases},
\end{align}
then using the fact $g(b) \geq e^{b/n}$ for any positive integer $n$, we have,
\begin{align}
      f(w,\delta E,x) \leq & \frac{c_{h,\hb}\boldsymbol c g(b)}{x^{2\Delta} \pi} \frac{\Gamma(\Delta)^2}{\Gamma(2\Delta)} \sum_{n=1}^\infty \frac{ w^n}{n^{2\Delta-1}}    e^{  \pi n \delta E - b n}, \nonumber  \\
      \leq & \frac{c_{h,\hb}\boldsymbol c g(b)}{x^{2\Delta} \pi} \frac{\Gamma(\Delta)^2}{\Gamma(2\Delta)}    \text{Li}_{2\Delta-1}\left(we^{  \pi  \delta E - b }\right).
\end{align}
Finally using the fact that $\lim_{z \rightarrow \infty}\text{Li}_\nu(e^z) \sim \log^\nu (z)$, 
\begin{align}
    \lim_{w\rightarrow \infty}\frac{f(w,\delta E,x)}{w} \sim \lim_{w \rightarrow \infty} \frac{1}{x^{2\Delta}} \frac{\left(\log(w)  + \pi \delta E - b\right)^{2\Delta -1}}{w} =0,
\end{align}
We find that $f(w,\delta E,x)/w$ vanishes at infinity and thus our analytic continuation (\ref{contf}) is valid for all $n\geq 1$. Note that $f(w,\delta E,x)/w$ is unbounded if $\delta E \rightarrow \infty$ or $x=0$. This is exactly consistent with our argument for (\ref{proof}).


\section{Higher $d$ Holographic CFT} \label{holo}
The purpose of this section is to provide evidence that the discontinuity of the Renyi transform provides the correct analytic continuation of Renyi correlators. We will do this in the context of large $N$ holographic CFT$_d$ for $d>2$. Although the Renyi transform is hard to evaluate for a generic replica correlator, we can make progress by considering the following replica correlator,
\begin{align}
    \tilde F_n(x) = \Tr \left( \rho^\frac{n}{2} O(x) \rho^\frac{n}{2} O(x) \right). \label{newf}
\end{align}
We will show that the discontinuity of the Renyi transform provides the correct analytic continuation,
\begin{align}
    \tilde F_n(x) = \frac{1}{2\pi i } \!\!\!\!\!\! \int\displaylimits_{w=e^{2\pi E_{min}}}^{\infty} \!\!\!\!\!\!\!\! dw  \, \, \frac{\disc \, \tilde f(w,x)}{w^{n+1}}. \label{newd}
\end{align}
We define the Renyi transform of the replica correlator $\tilde F_n(w,x)$ just like before as,
\begin{align}
    \tilde f(w,x)= \sum_{n=1}^\infty \tilde F_n(x) w^n.
\end{align}
Working with a correlator of the form (\ref{newf}) instead of (\ref{main}) has some obvious limitations. The most important being that there is no extra parameter like $s$ in (\ref{newf}) that would allow us to take an inverse integral transform. As a result, we do not have a relation of the form (\ref{discmeth}) that could allow us to obtain the individual off-diagonal elements of modular ETH from the discontinuity of the Renyi transform. 

As before we are interested in the $n$-sheeted cover of $\mathbb R_d$, branched at the boundary of half space at $x=0$,
\begin{align}
    ds^2 =  \frac{x^2}{R^2} d\theta^2 + dx^2+ dy_i^2 =  \frac{x^2}{R^2}  \left( d\theta^2 +  R^2 \frac{dx^2 + dy_i^2}{x^2}\right). \label{cfthyp}
\end{align}
This is conformally equivalent to $S^1_\beta \times \mathbb H_{d-1} $ with $\beta=2\pi n R$. In real time this corresponds to a conformal field theory at finite temperature, 
\begin{align}
    T=\frac{1}{2\pi n R}, \label{cftemp}
\end{align}
on the spatial manifold $\mathbb H_{d-1}$ with curvature radius $R$. Here $R$ is an arbitrary dimensionful parameter and has no physical significance by itself. Thus calculating the Renyi entropy of the half space reduces to calculating the thermal entropy on the Hyperbolic cylinder as was originally argued by Casini, Huerta and Myers \cite{Casini:2011kv,Hung:2011nu}. 

While calculating the thermal entropy in an arbitrary CFT is by itself a hard question, we can make progress for large $N$ CFTs using AdS/CFT. The thermal state on the boundary theory is dual to a black hole with the appropriate even horizon geometry \cite{Hung:2011nu}. In our case it corresponds to a Hyperbolic event horizon \cite{Emparan_1999},
\begin{align}
    ds^2 & = - \left( \frac{r^2}{\ell^2} - \frac{\mu}{r^{d-2}} - 1 \right) \frac{\ell^2}{R^2} dt^2 +  \frac{dr^2}{\left( \frac{r^2}{\ell^2} - \frac{\mu}{r^{d-2}} - 1 \right)} + r^2 \left(\frac{ dx^2 + \sum_{i=1}^{d-2}dy_i^2}{x^2} \right), \nonumber \\
   & \equiv - f(r) \frac{\ell^2}{R^2}  dt^2 +  \frac{dr^2}{f(r)} + r^2 dH_{d-1}^2. \label{hyp}
\end{align}
$\mu$ is related to the mass of the blackhole and can be fixed in terms of the event horizon radius by demanding $f(r_h)=0$ to obtain,
\begin{align}
    f(r) = \frac{r^2}{\ell^2} - 1 - \left(\frac{r_h}{r}\right)^{d-2} \left(\frac{r_h^2}{\ell^2}-1\right).
\end{align}
The extra factor of $\ell^2/R^2$ multiplying the time-time component is to ensure that (\ref{hyp}) is conformally equivalent to (\ref{cfthyp}). The temperature of the black hole can be found by the usual euclidean cigar trick,
\begin{align}
    T= \frac{d r_h^2 - (d-2) \ell^2}{4\pi \ell R r_h} \label{temp},
\end{align}
Using this and (\ref{cftemp}) we have,
\begin{align}
    \frac{r_h}{\ell} = \frac{1 + \sqrt{1- 2d n^2 + d^2 n^2}}{n d}. \label{rhn}
\end{align}
This relationship will be useful later when we take the Renyi transform. The thermal entropy is just given by the black hole entropy. 

We want to calculate the replica correlator (\ref{newf}) on this blackhole background (\ref{hyp}). In a large $N$ holographic CFT this involves solving the bulk Klein-Gordon equation in the given background. This is hard to do in general, and as is usual we will focus on the case where $N \gg \Delta_{1,2} \gg 1$. In this limit, we can use the geodesic approximation to calculate the two point function,
\begin{align}
    \braket{O(x_1,t_1) O(x_2,t_2)} = \text{exp} \left(-\Delta \mathcal L \right), \label{ads}
\end{align}
where $\Delta$ is the conformal dimension of the scalar operator $O(x_i,t_i)$ and $\mathcal L$ is the proper length of the bulk geodesic connecting the two points. 
We will follow the prescription of Faulkner et al. \cite{Faulkner:2018faa} where it was argued that the correlator of the form (\ref{newf}) can be obtained by computing the length of the geodesic that extends from the boundary to the bulk $Z_N$ fixed point -- the horizon. 

The four velocity of a particle moving on a geodesic is given by,
\begin{align}
    u^\mu u_\mu = - \kappa,
\end{align}
where $\kappa=-1,0,1$ for timelike, null and spacelike geodesics respectively. In the blackhole background of (\ref{hyp}) this becomes,
\begin{align}
    -f(r) \left( \frac{dt}{d\tau} \right)^2 + \frac{1}{f(r)} \left( \frac{dr}{d\tau} \right)^2 + \frac{r^2}{x^2} \left( \frac{dx}{d\tau}\right)^2  + \frac{r^2}{x^2} \left( \frac{d y_i}{d\tau}  \right)^2  = \kappa. \label{geo}
\end{align}
Using the isometries of the background
we can write the following conserved quantities,
\begin{align}
     E & = - g_{\mu \nu} \frac{dx^\mu}{d\tau} \left( \frac{\partial }{\partial t} \right)^\nu = f(r) \dot t, \nonumber \\ 
     L_i & = g_{\mu \nu} \frac{dx^\mu}{d\tau} \left( \frac{\partial }{\partial y^i} \right)^\nu = \frac{r^2}{x^2} \dot{y}^i, \nonumber  \\ 
     D & = g_{\mu \nu} \frac{dx^\mu}{d\tau} \left(x\frac{\partial}{\partial x}+ y^i\frac{\partial}{\partial y^i} \right)^\nu= \frac{r^2}{x^2} \left( x \dot{x}  + y_i  \dot{y}^i \right), \label{conv}
\end{align}
where dot refers to the derivative with respect to proper time. We are interested in a geodesic that starts from the boundary and travels all the way to the horizon radially. This corresponds to setting $E=D=L_i=0$.\footnote{ If we were to calculate geodesic lengths for a generic replica correlator like (\ref{main}) with time like separation say $\Delta s$, the conserved charge $E$ would be nonzero. The proper length is still given by an integral of the form (\ref{pint}), and can be actually computed in the Hyperbolic black hole background. However the proper length is obtained as a function of the conserved charge $E$ and not $\Delta s$. The time separation $\Delta s$ is obtained by integrating $E= f(r) \dot t$. This can be also done in practice to obtain $\Delta s = g(E,r_h)$ where $g$ is an extremely complicated function that we know explicitly. However, it is almost an impossible problem to invert this function. This is a common issue in calculating geodesics lengths in the bulk \cite{Fidkowski:2003nf}. Evaluating the Renyi transform precisely requires this inversion. This is why computing the Renyi transform of the replica correlator (\ref{main}) is hard. \label{foot}} Setting $\kappa=1$ in (\ref{geo}) we obtain,
\begin{align}
      \dot r^2 = f(r). \label{turn}
\end{align}
The proper length of the geodesic is then given by,
\begin{align}
    \mathcal L & = \int i \frac{d\tau}{dr} dr = 2 i \int_{r_h}^\Lambda 
    \frac{dr}{\dot r} = 2  \int_{r_*}^\Lambda \frac{dr}{\sqrt{E^2 - f(r)}}, \label{pint} \\
     & =  \frac 1 2 \log \left(\frac{16 \Lambda ^4}{\left(1-2 r_h^2\right)^2} \right), \label{tau}
\end{align}
where we have set $E=0$.
We can now evaluate the replica correlator $\tilde F_n(x)$. Using the holographic dictionary for correlators (\ref{ads}) and the expression (\ref{rhn}) for the event horizon radius $r_h$ in terms of the Renyi parameter $n$, we obtain, after holographic renormalisation,
\begin{align}
    \tilde F_n(x) = \left( \frac{\sqrt{8 n^2+1}+1}{4 n^2}  \right)^{\Delta}. \label{fexp}
\end{align}
In the rest of the section we will explicitly evaluate the Renyi transform and show that our analytic continuation (\ref{newd}) involving the discontinuity of the Renyi transform indeed reproduces the replica correlator (\ref{fexp}). This will serve as a non-trivial check of our formula.

The Renyi transform is given by the infinite sum, 
\begin{align}
    \tilde f(w) & = \sum_{n=1}^\infty \tilde F_n(x) w^n =  \sum_{n=1}^{\infty} \left( \frac{\sqrt{8 n^2+1}+1}{4 n^2}  \right)^{\Delta} w^n. \label{nfir}
\end{align}
Note that the discontinuity of each individual term in the sum vanishes since $n$ is an integer. This is a common theme that occurs in dispersion relations involving analytic functions \cite{Caron-Huot:2017vep}. It seems that we are free to change the individual coefficients i.e. $\tilde F_n(x)$ in the $w^n$ expansion without changing the discontinuity. However the point is that since $\tilde f(w)/w$ is bounded at infinity, the individual coefficients $\tilde F_n$ are tightly constrained. In fact the existence of our analytic continuation (\ref{newd}) shows that these coefficients form an analytic function in $n$. To proceed further, we thus need to evaluate the infinite sum first. Using the multinomial theorem we have,
\begin{align}
    \tilde f(w) &  = \sum_{n=1}^{\infty} \sum_{m,p=0}^{\infty}   \binom{\Delta }{m} \binom{\frac{m}{2}}{p} n^{m-2p-2\Delta} 2^{-2 \Delta +\frac{3 m}{2}-3 p} w^n, \nonumber \\
    & =  \sum_{m,p=0}^{\infty}   \binom{\Delta }{m} \binom{\frac{m}{2}}{p} 2^{-2 \Delta +\frac{3 m}{2}-3 p} \text{Li}_{2p+2\Delta-m}(w).
\end{align}
In the last line we used the series expansion of the polylogarithm. Using the discontinuity of the polylogarithm (\ref{li}) we have,
\begin{align}
    \disc \tilde f(w) & = 2\pi i \! \! \sum_{m,p=0}^{\infty} \binom{\Delta }{m} \binom{\frac{m}{2}}{p} \frac{  2^{-2 \Delta +\frac{3 m}{2}-3 p} \log ^{2 \Delta +2p-m-1}(w)}{\Gamma (2 p+2 \Delta - m)}, \nonumber \\
    & = 2\pi i \! \hspace{0.055cm} \sum_{m=0}^{\infty} \binom{\Delta }{m} \frac{2^{\frac{3 m}{2}-2 \Delta }  \log ^{2 \Delta -m-1}(w) \, _1F_2\left(-\frac{m}{2};\Delta -\frac{m}{2},\Delta -\frac{m}{2} +\frac{1}{2};-\frac{\log ^2(w)}{32} \right)}{\Gamma (2 \Delta -m)}.
\end{align}
Using this result for the disc in our analytic continuation (\ref{newd}) we finally obtain,
\begin{align}
    \tilde F_n(x) & = \frac{1}{2\pi i}\int_{1}^\infty dw \frac{\disc \tilde f(w,x)}{w^{n+1}}  = \sum_{m=0}^{\infty} 2^{\frac{3 m}{2}-2 \Delta } \left(\frac{1}{8 n^2}+1\right)^{m/2} \binom{\Delta }{m} n^{m-2 \Delta }, \nonumber \\
    & = \left(\frac{\sqrt{1+8n^2}+1}{4n^2}\right)^{\Delta }.
\end{align}
This exactly matches with (\ref{fexp}) and provides a non-trivial consistency check of our formula (\ref{newd}). If the Renyi transform (\ref{nfir}) that we calculated here explicitly, did not have the kind of analytic structure that we argued in section \ref{prop}, our analytic continuation (\ref{newd}) would not have reproduced the replica correlator (\ref{fexp}) that we started with originally.


\section{Discussion} \label{discu}
In this note we defined a new object, the Renyi transform $f(w,s,x)$, for certain replica correlators $F_n(s,x)$. We argued that the discontinuity of the Renyi transform in the complex $w$ plane provides the unique analytic continuation (\ref{contf}) of $F_n(s,x)$ in $n$, as long as $f(w,s,x)$ satisfies certain boundedness properties at infinity in the complex $w$ plane. We showed that $f(w,s,x)$, for nonzero $s$ and $x$, satisfies those properties if and only if the density of states $\mu(E)$ and matrix elements $\braket{E|O(x)|E+\delta E}$ grow sub-exponentially in the modular energy $E$. In this process, we also discovered a non-trivial identity (\ref{discmeth}) that relates the discontinuity of the Renyi transform with the off-diagonal elements of modular ETH.

It would be extremely interesting if we could evaluate the Renyi transform of replica correlators in quantum field theories with spacetime dimension $d>2$. The discontinuity of the Renyi transform would then provide us the extremely non-trivial matrix elements $\braket{E|O(x)|E+\delta E}$ via the relation (\ref{discmeth}). In particular, for conformal field theories as discussed near (\ref{hhe}), this would provide us the $\text{Heavy-Light-}\widetilde{\text{Heavy}}$ three point OPE coefficients in higher dimensions on hyperbolic space. In the case of large $N$ holographic theories, these OPE coefficients \cite{Fitzpatrick:2015zha} are directly related to thermalisation of hyperbolic blackholes. In such cases we could use the geodesic approximation to calculate the Renyi transform as done in section \ref{holo}, but for reasons mentioned in footnote \ref{foot} this is difficult to do for generic replica correlators. We hope to return to this question in the future.

On the other hand it would be also interesting if the Renyi transform could be evaluated independently. 
Our analytic continuation (\ref{contf}) would then provide an alternate method of calculating replica correlators. As mentioned in the introduction such correlators are extremely useful when studying entanglement entropy of excited states and/or shape deformations. Morally a similar feat has been accomplished for Renyi entropies in $AdS_2$ by \cite{Penington:2019kki}. They define the resolvent for the density matrix which is quite similar to our Renyi transform, and compute the Renyi entropies by first evaluating the resolvent of the density matrix.

Our work can be easily generalised to higher point replica correlators such as,
\begin{align}
    F_n(s_i,x_i) = \Tr \left(\rho^n  O(\vec x_1) \rho^{\frac{is_2}{2\pi}} O(\vec x_2) \rho^{-\frac{is_2}{2\pi}} \rho^{\frac{is_3}{2\pi}} O(\vec x_3) \rho^{-\frac{is_3}{2\pi}} \ldots \right).
\end{align}
We can define the Renyi transform of this correlator just like before and show that the discontinuity provides the unique analytic continuation for a higher point replica correlator as long as the density of states and the matrix elements $\braket{E|O|E + \delta E }$ grow sub-exponentially in the modular energy $E$. Obviously the higher point replica correlators and Renyi transform are much harder to compute than the two point case, but we can make progress in CFTs that allow large $N$ factorisation \cite{Belin:2018juv}. It would also be extremely interesting to extend our work to replica correlators with multiple modular flows. Such correlators have been very useful recently for e.g. in deriving the modular chaos bound \cite{Faulkner:2018faa}, and we hope to report on it in the future.

\acknowledgments
It is a pleasure to thank Alexandre Belin, Matthew Dodelson, Tom Faulkner, Shamit Kachru, Edward Mazenc and Tomonori Ugajin for several interesting discussions and comments on the draft. I would like to especially thank Onkar Parrikar for comments on the draft, and for collaboration during the early stages of this project. Finally, I would like to thank Raghu Mahajan for several interesting discussions, comments on the draft and encouragement throughout the duration of this work. This work has been supported by the Department of Energy grant BES DE-SC0020007 and the Simons collaboration on Ultra Quantum Matter.





 \bibliographystyle{JHEP}
 \bibliography{biblio}

\end{document}